\documentclass[%
 reprint,
 amsmath,amssymb,
 aps,
nofootinbib,
]{revtex4-2}
\usepackage{graphicx}
\usepackage{epstopdf}
\usepackage{color}
\usepackage{multirow}
\usepackage{subeqnarray}
\usepackage{setspace}
\usepackage{palatino}
\usepackage{dcolumn}
\usepackage{bm}
\usepackage[utf8]{inputenc}
\usepackage[T1]{fontenc}
\usepackage{mathptmx}
\usepackage{amsmath,amsfonts,amscd,amssymb,amsthm}
\usepackage{hyperref}
\usepackage{commath}

\usepackage{algorithm,algcompatible,algorithmicx}
\usepackage{algpseudocode}

\algnewcommand\algorithmicinput{\textbf{Input:}}
\algnewcommand\INPUT{\item[\algorithmicinput]}
\algnewcommand\algorithmicoutput{\textbf{Output:}}
\algnewcommand\OUTPUT{\item[\algorithmicoutput]}
\algnewcommand\algorithmicoptional{\textbf{Optional:}}
\algnewcommand\OPTIONAL{\item[\algorithmicoptional]}
\usepackage{overpic}
\usepackage{cancel}
\usepackage{rotating}
\usepackage{url}
\usepackage{caption}
\usepackage{subcaption}
\usepackage{mathtools}

\DeclareMathOperator*{\argmin}{arg\,min}
\usepackage{setspace}
\usepackage[bottom,flushmargin,hang,multiple,symbol]{footmisc}

\usepackage[normalem]{ulem}

\begin{document}

\title{Topology optimization for inverse magnetostatics as sparse regression:\\application to electromagnetic coils for stellarators}
\author{Alan A. Kaptanoglu}\thanks{Corresponding author (akaptano@umd.edu).}
 \affiliation{Institute for Research in
Electronics and Applied Physics,\\ University of Maryland, College Park, MD, 20742, USA\looseness=-1} 
\author{
Gabriel P. Langlois}
 \affiliation{Courant Institute of Mathematical Sciences, New York University, New York, NY, 10012, USA\looseness=-1} 
\author{Matt Landreman}
 \affiliation{Institute for Research in
Electronics and Applied Physics,\\ University of Maryland, College Park, MD, 20742, USA \looseness=-1} 

\begin{abstract}
Topology optimization, a technique to determine where material should be placed within a predefined volume in order to minimize a physical objective, is used across a wide range of scientific fields and applications. 
A general application for topology optimization is inverse magnetostatics; a desired magnetic field is prescribed, and a distribution of steady currents is computed to produce that target field.
In the present work, electromagnetic coils are designed by magnetostatic topology optimization, using volume elements (voxels) of electric current, constrained so the current is divergence-free.
Compared to standard electromagnet shape optimization, our method has the advantage that the nonlinearity in the Biot-Savart law with respect to position is avoided, enabling convex cost functions and a useful reformulation of topology optimization as sparse regression. 
To demonstrate, we consider the application of designing electromagnetic coils for a class of plasma experiments known as stellarators.
We produce topologically-exotic coils for several new stellarator designs and show that these solutions can be interpolated into a filamentary representation and then further optimized.
\end{abstract}

\maketitle

\section{Introduction} \label{sec:intro}
Topology optimization aims at solving a fundamental engineering problem; where should material be placed in a predefined volume in order to minimize some physical objective function? This general problem spans a wide range of scientific disciplines and a large number of approaches have been developed for carrying out variations of topology optimization~\cite{sigmund2013topology}. General topology optimization can be written
\begin{align}
    \label{eq:general_topology_opt}
    \min_{\bm \alpha} &f(\bm \alpha),\\ \notag    s.t. \quad &\mathcal{\bm{C}}_0(\bm\alpha) = 0, \\ \notag
  \quad &\mathcal{\bm{C}}_1(\bm\alpha) \leq 0, \\ \notag
  &\alpha_i = 0\text{ or }1, \forall i,
\end{align}
where the elements of $\bm \alpha$  are the optimizable degrees of freedom with permissible values of only 0 or 1, $\mathcal{C}_0$ and $\mathcal{C}_1$ are general constraints on $\bm \alpha$, and $f$ is the primary objective.
This binary, constrained, and general form of the problem is very challenging. However, many problems, including those addressed in this work, have a convex objective and convex constraints. This assumption can make high-dimensional topology optimization much more tractable, although the nonconvexity from the binary nature of the problem remains.

Traditional density-based approaches relax the binary problem to a continuous one with $0 \leq \alpha_i \leq 1$, with an additional penalty for values of $\alpha_i$ between 0 and 1~\cite{bendsoe1989optimal,allaire1993numerical,stolpe2001alternative}. This density approach has been used extensively in structural engineering as well as for designing permanent magnets for a class of plasma experiments called stellarators~\cite{zhu2020topology,hammond2020geometric,qian2021stellarator,qian2022simpler}. 
There are many other approaches to topology optimization and we refer the reader to the review in Sigmund and Maute~\cite{sigmund2013topology}. As far as we are aware, until the present work topology optimization has not been performed by solving a continuous version of Eq.~\eqref{eq:general_topology_opt} with the $l_0(\bm \alpha) = \|\bm \alpha\|_0$ pseudo-norm, an operator that counts the number of nonzero elements in $\bm\alpha$. Use of the $l_0$ norm turns the problem into a form of sparse regression. Presumably, this approach has not been favored because it presents a nonconvex, nonsmooth loss term to optimize, preventing the application of traditional gradient or Hessian-based solvers. Nonetheless, this problem can be solved effectively in some important applications, e.g. as we will show in the present work, for designing electromagnetic coils.
Before describing this formulation and why it is advantageous, particularly in electromagnetic coil design, we review our motivating application from the field of plasma physics.

\subsection{Stellarator optimization}\label{sec:stell_opt_intro}
The design of electromagnetic coils is required in a large number of scientific and engineering domains.
In one common situation, considered here, a target magnetic field in some volume is given, and the goal is to find a configuration of magnets outside that volume to produce the desired field.
Examples of this problem are producing uniform fields and uniform field gradients for magnetic resonance imaging \cite{hidalgo2010theory, chen2017electromagnetic}, and producing uniform dipole or quadrupole fields for the beam optics in particle accelerators \cite{russenschuck2011field}.
This problem is an ill-posed inverse problem because many different magnet designs can produce a nearly identical target magnetic field via the Biot-Savart law. 

This inverse magnetostatics problem is also critical for stellarators, a class of plasma devices commonly considered for future nuclear fusion reactors.
Stellarator design relies on sophisticated coil optimization algorithms in order to produce ideal magnetic fields for confining plasma~\cite{grieger1992physics,imbert2019introduction}. These three-dimensional magnetic fields must be carefully shaped in order to provide high-quality confinement of charged particle trajectories and many other physics objectives. 

Optimizing stellarators is typically performed in two stages. The first is a configuration optimization using fixed-boundary magnetohydrodynamic equilibrium codes to obtain plasma equilibria with desirable physics properties~\cite{nuhrenberg1988quasi,spong1998j,drevlak2018optimisation,landreman2021simsopt,dudt2023desc}. Important metrics to minimize for nuclear fusion devices include deviations from quasi-symmetry (an unusual symmetry in magnetic fields that enables particle confinement), fast ion losses, and magnetohydrodynamic instability~\cite{hegna2022improving}. 

After obtaining the optimal magnetic field in this first stage, magnets must be designed to produce these fields, subject to a number of engineering constraints such as a minimum coil-to-coil distance, maximum forces on the coils~\cite{robin2022minimization}, maximum curvature on the coils~\cite{zhu2017new}, and many other requirements. There has also been recent work combining the two optimization stages into a single overall optimization~\cite{henneberg2021combined, giuliani2022single, jorge2023single}.
In any case, the result is that stellarator coils are often complex three-dimensional shapes, raising the cost and difficulty of manufacturing. A primary cost driver of the W-7X and NCSX stellarator programs was the manufacture and assembly of complex coils with tight engineering tolerances~\cite{erckmann1997w7,strykowsky2009engineering}. 

\subsection{Coil optimization}\label{sec:coil_opt_intro}
Formulation of coil design in the language of topology optimization, and later, sparse regression, facilitates the effective use of a large literature from across scientific disciplines. 
To motivate our eventual formulation, consider first a general situation in which there is a surface $S'$ (or volume $V'$) of current sources and a surface $S$ (or volume $V$) where we want to match a target magnetic field $\bm B$ (or its magnitude $B$, or just one of its components). A simple example is the following inverse magnetostatic optimization problem,
\begin{align}
\label{eq:simple_inverse_magnetostatic}
    \min_{\bm J}\int_S &\| \bm B_\text{coil} - \bm B_\text{target}\|^2d\bm r,
\end{align}
where $\bm J$ are a set of coil current densities, $\bm B_\text{coil}(\bm r)$ is the magnetic field generated by the coils, $\bm B_\text{target}(\bm r)$ is the magnetic field desired on the surface $S$, and $\bm r$ is a coordinate vector. Eq.~\eqref{eq:simple_inverse_magnetostatic} represents the general situation in which a set of coils and magnets are desired that match a target magnetic field. Notice that we have assumed a set of available and fixed spatial locations for the coils because only the coil currents are used as optimization variables.

Traditionally, coil design for stellarators is performed with either the \textit{winding surface} or \textit{filament} method. The winding surface optimization problem~\cite{landreman2017improved} is a variation of Eq.~\eqref{eq:simple_inverse_magnetostatic}. The goal is to minimize the normal component of $\bm B$ on the surface of a plasma $S$, and the sources lie on a winding surface $S'$ that is pre-defined by the practitioner. $S'$ is typically prescribed by extending the plasma boundary outward using an overall offset multiplied by the normal vectors on this surface; see Appendix~\ref{sec:appendix_grid_generation} for additional details. Now the following optimization problem is solved:
\begin{align}
\label{eq:winding_surface_optimization}
    \min_{\bm J}\int_S &\|(\bm B_\text{coil} - \bm B_\text{target})\cdot\hat{\bm n}\|^2d\bm r + \kappa\int_{S'} \|\bm J(\bm r')\|^2d \bm r', 
    \\ 
    \label{eq:biot_savart}
    \bm B_\text{coil}(\bm r) &\equiv \frac{\mu_0}{4 \pi}\int_{S'}\frac{\bm J(\bm r') \times (\bm r - \bm r')}{\|\bm r - \bm r'\|^3}d\bm r'.
\end{align}
Throughout this work, quantities associated with the coil surface (and later, volume) are denoted with a prime, and norms without subscripts, $\|\cdot\|$, indicate vector magnitudes. Notice that the current density $\bm J$ in Equations~\eqref{eq:winding_surface_optimization} and ~\eqref{eq:biot_savart} is a surface current density in Amperes per meter because we have assumed for now that the sources lie entirely on a surface $S'$; the coils are represented by a continuous sheet current on the winding surface. Here, $\bm r'$ is a source position, $\hat{\bm n}$ is the plasma unit normal vector ($\bm n$ for a non-unit normal vector), $\mu_0$ is the vacuum permeability, and $\kappa$ is a scalar hyperparameter that determines how strongly to penalize large and potentially unrealistic currents.
The $\bm B_\text{target}\cdot\hat{\bm n}$ term can represent normal magnetic field contributions from other sources, including other coils or magnets, or contributions from finite plasma current. The Tikhonov regularization term proportional to $\kappa$ is traditionally used to deal with the ill-posedness intrinsic to coil optimization; without additional optimization criteria, very different coil sets can produce similar residuals in $\bm B \cdot \hat{\bm n}$ on the plasma surface. In winding surface optimization without Tikhonov regularization, large and unrealistic surface currents can be generated. These currents can then overfit to the quadrature points on the plasma surface that were used to discretize the first integral in Eq.~\eqref{eq:winding_surface_optimization}.

A global, smooth, and periodic Fourier basis is used for the currents in winding surface optimization. This representation results in a linear least-squares problem that can be easily solved, 
and produces surface currents that are a priori continuous and divergence-free. However, in principle we could make different assumptions about the spatial variation and topology of the currents by expanding the currents locally, using local spatial basis functions. 
The advantage of the winding surface method over the filament-based algorithms is that the surface is defined before optimization, and subsequently avoids a much more complicated optimization over spatial degrees of freedom. This feature is also a disadvantage, since the predefined and fixed spatial grid is a strong constraint on the space of possible coil shapes. 

Conversely, the filamentary method allows for complex spatial dependence by representing the coils as zero-thickness curves in three-dimensional space and optimizing the spatial degrees of freedom of the curves. However, this approach leads to a significantly more complicated optimization problem. From an optimization standpoint, it is critical to notice that the Biot-Savart law in Eq.~\eqref{eq:biot_savart}, regardless if the coils are represented by one-dimensional curves or three-dimensional volumes, is linear in $\bm J$ but nonlinear in $\bm r'$.
Moreover, most of the additional engineering constraints for coils, such as minimum coil-coil distances, are also nonlinear functions of $\bm r'$. 
These nonlinearities guarantee that filament optimization is highly nonconvex, whereas the winding surface method is convex and much simpler to solve. Our new method aims to combine the best features of both: freedom in all three spatial dimensions like filaments, but with the convexity and linearity of the winding surface method. 

\subsection{Contributions of this work}
\label{sec:contributions}
In the present work, we provide an algorithm that can be used to solve topology optimization problems of the form of Eq.~\eqref{eq:general_topology_opt}, rewritten
\begin{align}
    \label{eq:general_sparse_regression}
    \min_{\bm \alpha} &\left\{f(\bm \alpha) + \lambda\|\bm \alpha \|_0\right\},\\ \notag    s.t. \quad &\mathcal{\bm{C}}_0(\bm\alpha) = 0, \quad
  \quad \mathcal{\bm{C}}_1(\bm\alpha) \leq 0,
\end{align}
with the additional assumption that the $\lambda=0$ subproblem can be solved with reasonable computational efficiency. This assumption is  certainly true for a large class of convex objectives and convex constraints. Note also that Eq.~\eqref{eq:general_sparse_regression} is not quite equivalent to the binary problem, 
but if necessary, upper bounds can be prescribed in the form of linear constraints on the $\alpha_i$ so that the elements of $\bm\alpha$ take only two possible values. A large volume of literature exists for solving a relaxation of Eq.~\eqref{eq:general_sparse_regression} with the $l_0$ norm replaced with the $l_1$ norm~\cite{Tibshirani1996lasso,bertsimas2020sparse}, since then Eq.~\eqref{eq:general_sparse_regression} is convex and can be easily solved. 
However, when a parameter is nonzero in the $l_1$ problem, it can take any value. This is unsuitable for our task since it is important in coil design that the currents in the voxels are either very large or zero, approximating the binary structure of the problem.

Next, we use this new sparse regression formulation of topology optimization to generate coil designs without resorting to winding surfaces (only surface currents can exist) or filaments (zero-thickness curves). This is the first demonstration of stellarator coil design using topology optimization. We further illustrate our new method by generating a series of coil sets for three recent high-performance stellarators. Because the currents are local and vary throughout a volume, we  refer to our optimization technique as the current voxel method. 

Unlike the traditional methods, after a coil volume is defined, the coil shape topology is an output of the optimization rather than an input by the user. This is an important step because the optimal coil topology for a particular stellarator is often unclear, so researchers often manually try a number of possible configurations. In this sense, our current voxel optimization can also be seen as providing a principled initial topology and set of coils for further optimization using other coil optimization routines. To demonstrate this use case, we take two helical coil designs generated by our new method and initialize filament optimizations, which perform further solution polishing.
Lastly, the methodology described here is implemented in the open-source SIMSOPT code~\cite{landreman2021simsopt,SimsoptURL},
which was used to generate the results in the present paper. 

\section{Current voxel optimization}\label{sec:methodology}
So far, we have outlined that our new coil optimization should generate coils that vary in three spatial dimensions but avoid the Biot-Savart nonlinearity in $\bm r'$. We now show that these requirements produce an optimization problem equivalent to the topology optimization in Eq.~\eqref{eq:general_sparse_regression}.

Consider the following variation of the winding surface objective in Eq.~\eqref{eq:winding_surface_optimization}, in which the current is now allowed to vary continuously within some \textit{volume} $V'$ surrounding the plasma surface.
Since the coil volume is predefined, there is no shape optimization, a feature in common with the original winding surface method. 
For simplicity, let us assume that we can reasonably decompose $V'$ (the ``winding volume'') into a three-dimensional mesh of grid cells, e.g., rectangular cubes, which we refer to as the current voxels. 
Then we have $D$ discrete but continuously connected, rectangular grid cells with volumes $V_k'$ such that $\cup_{k=1}^DV_k' \approx V'$ and some amount of current in each cell,
\begin{align}
   \label{eq:Bnormal_new}
    \bm B_\text{coil}(\bm r)\cdot\hat{\bm n} = -\frac{\mu_0}{4 \pi}\sum_{k=1}^D\int_{V_k'}\frac{\hat{\bm n}\times (\bm r - \bm r_k')}{\|\bm r - \bm r_k'\|^3}\cdot\bm J_{k}(\bm r_k') d\bm r_k'.
\end{align}
If there are no existing coils or applied fields, i.e. $\bm B_\text{target} = 0$, then the trivial solution $\bm B_\text{coil} = 0$ needs to be avoided in the optimization. 
There are numerous strategies for preventing the trivial solution. One strategy in filamentary coil optimization 
is to set a nonzero current in one of the filaments. Another strategy to avoid the trivial solution specifies the toroidal flux in a poloidal cross section~\cite{zhu2017new}. Here, we consider instead fixing a target current value, $I_\text{target}$, by computing a line integral 
\begin{align}  \label{eq:Itarget}
\mu_0 I_\text{target} = \oint_\gamma\left(\bm B_\text{coil} - \bm B_0\right) \cdot \bm{dl},
\end{align}
around a toroidal loop $\gamma$ that is on or in the plasma, e.g., the magnetic axis or the plasma boundary at $\theta = 0$. For all of the examples illustrated in this work, we use the latter. $\bm B_0$ is the magnetic field along the toroidal loop from other sources, e.g. finite plasma current.
Equation~\eqref{eq:Itarget} requires only the computation of a single integral, and it is explicitly shown in Appendix~\ref{sec:appendix_formalism} to be linear in the optimization variables defined in the next section. Typically, the solution need not \textit{exactly} match the target current value, so we incorporate the squared residual of~\eqref{eq:Itarget} as another linear least-squares term in the optimization problem that will be described shortly.

\subsection{A finite element basis for the currents}\label{sec:Galerkin_expansion}
In each cell, $\bm J_k$ in Eq.~\eqref{eq:Bnormal_new} must necessarily have nontrivial spatial dependence for nontrivial coil designs.
The local spatial variation comes from expanding each $\bm J_k$ in a finite element basis, and the coefficients of that basis will later become the variables for optimization, 
\begin{align}
\label{eq:j_expansion}
    \bm J_k \equiv \bm{\alpha}_k\cdot\bm{\phi}_k(\bm r_k') =\sum_{i=1}^N\alpha_{ik}\bm\phi_{ik}. 
\end{align}
The $\bm \phi_i$ represent a chosen set of spatial basis functions and we use the divergence-free basis of linear polynomial vectors as in Cockburn~\cite{cockburn2012discontinuous}, so that the $\bm J_k$ are divergence-free in each grid cell. 
Of course, higher-order polynomial basis functions can be used for improved convergence.

However, as of now there can still be deviations from global current conservation, since there can be flux jumps across cell interfaces, and the current is not imposed to be continuous. To solve the former problem, we can impose that surface-averaged flux jumps across cells vanish:
\begin{align}
\label{eq:flux_jumps_vanish}
\int_{V_k' \cap V_l'} &\hat{\bm n}'\cdot\left[\bm J_k(\bm r_k') - \bm J_l(\bm r_k')\right]d^2 r_k' = 0.
\end{align}
This constitutes at most six linear constraints per cell on the $\bm \alpha_k$, for a total of $N_c < 6D$ constraints. On average, we expect $N_c \sim 3D$, since adjacent cells need only one constraint for their mutual interface, but the exact number will vary with the geometry of the voxel grid $V'$. 

For concreteness, the basis for degree-1, divergence-free polynomial vectors in three-dimensions can be chosen as (taking centers of the cell at $(x_k, y_k, z_k)$):
\begin{align}\label{eq:linear_polynomials}
&X_k \equiv \frac{x - x_k}{\Delta x_k}, \quad Y_k \equiv \frac{y - y_k}{\Delta y_k}, \quad Z_k \equiv \frac{z - z_k}{\Delta z_k}, 
\end{align}
\vspace{-0.2in}
\medmuskip=-3mu
\thinmuskip=-3mu
\thickmuskip=-3mu
\begin{align} \notag
    &\begin{bmatrix}
    1 \\
    0 \\
    0 \\
    \end{bmatrix},     \begin{bmatrix}
    0 \\
    1 \\
    0 \\
    \end{bmatrix},     \begin{bmatrix}
    0 \\
    0 \\
    1 \\
    \end{bmatrix},     \begin{bmatrix}
    Y_k \\
    0 \\
    0 \\
    \end{bmatrix},     \begin{bmatrix}
    Z_k \\
    0 \\
    0 \\
    \end{bmatrix},     \begin{bmatrix}
    0 \\
    0 \\
    X_k \\
    \end{bmatrix},     \begin{bmatrix}
    0 \\
    0 \\
    Y_k \\
    \end{bmatrix},     \begin{bmatrix}
    0 \\
    Z_k \\
    0 \\
    \end{bmatrix},     \begin{bmatrix}
    0 \\
    X_k \\
    0 \\
    \end{bmatrix},     \begin{bmatrix}
    X_k \\
    -Y_k \\
    0 \\
    \end{bmatrix},     \begin{bmatrix}
    X_k \\
    0 \\
    -Z_k \\
    \end{bmatrix}.
\end{align}
\medmuskip=4mu
\thinmuskip=3mu
\thickmuskip=5mu
In order to avoid \textit{local} jumps in $\nabla\cdot\bm J$ along cell interfaces (the constraints only guarantee that $\nabla\cdot\bm J = 0$ in an integral sense over the cell interface), 
we enforce that the $J_i$ component is continuous on the cell interface with normal vector $\hat{\bm n}' = \hat{x}_i$. It turns out that this can be entirely enforced by simply reducing the number of basis functions to five,
\begin{align}
    &\begin{bmatrix}
    1 \\
    0 \\
    0 \\
    \end{bmatrix},     \begin{bmatrix}
    0 \\
    1 \\
    0 \\
    \end{bmatrix},     \begin{bmatrix}
    0 \\
    0 \\
    1 \\
    \end{bmatrix},    
    \begin{bmatrix}
    X_k \\
    -Y_k \\
    0 \\
    \end{bmatrix},    \begin{bmatrix}
    X_k \\
    0 \\
    -Z_k \\
    \end{bmatrix}.
\end{align}
To summarize, we now have a representation that can produce discontinuous currents and the divergence-free property of $\bm J$ is everywhere satisfied.
More sophisticated finite element geometries and basis representations are a clear place for future improvements. For instance, an orthonormal, hierarchical, and high-numerical-precision basis of divergence free polynomials can also be constructed to arbitrary degree and dimension in tetrahedral domains~\cite{anantharamu2022arnoldi}.

\subsection{Finalizing the optimization problem}\label{sec:optimization_formulation}
We now have a useful spatial basis to represent the current density in each cell. 
The Biot-Savart calculation for the normal component of $\bm B_\text{coil}$ reduces to a simple matrix-vector product between the optimization variables $\bm \alpha \in \mathbb{R}^{ND}$ and a matrix $\bm A \in \mathbb{R}^{n_\theta n_\zeta\times ND}$, which can be computed once before optimization begins. Here $n_\theta$ and $n_\zeta$ represent the number of poloidal and toroidal quadrature points on the plasma surface, respectively.
The optimization to solve so far can be shown to be a linear least-squares problem in $\bm \alpha$ with linear equality constraints,
\begin{align}
    \bm C \bm \alpha = \bm 0.
\end{align}
For a degree-1 basis, there are $5D$ variables in $\bm \alpha = [\bm \alpha_1, ..., \bm \alpha_D]$, which is in principle enough free parameters to satisfy the $N_c$ linear constraints coming from the flux jump constraints. Note that $\bm C \in \mathbb{R}^{N_c \times ND}$, which can be large (since D can be size $\sim 10^4-10^5$) but tractable because it is very sparse. 

The last ingredient for our new optimization is crucial. As it stands, there will be nonzero current contributions in every cell in the prescribed coil volume, which will clearly not generate isolated, discrete coils of the type desired for stellarators. However, we can alter the optimization problem to contain an additional term, called the \textit{non-overlapping group} $l_0$ norm, $\lambda\|\bm \alpha\|_0^G$. 
The quantity $\|\bm \alpha\|_0^G$ is defined to be the number of cells for which $\alpha_{ik} = 0$ for all $i$ in the cell indexed by $k$. Thus, $\|\bm \alpha\|_0^G$ is reduced only when cell currents are fully zeroed out in a given voxel.
Including this term in the optimization will produce a set of sparse coils. 
In total, we show in Appendix~\ref{sec:appendix_formalism} how the optimization can be formulated as:
\begin{align}
\label{eq:final_opt}
    \min_{\bm \alpha}&\left\{f_B(\alpha) + \kappa f_K(\alpha) + \sigma f_I(\alpha) + \lambda\|\bm \alpha\|_0^G\right\},\\ \notag 
    s.t. \quad &\bm C\bm \alpha = \bm 0, \quad f_B(\alpha) \equiv \frac{1}{2}\|\bm A\bm \alpha - \bm b\|_2^2,
    \\ \notag 
    \quad f_K(\alpha) &\equiv \frac{1}{2D}\|\bm \alpha\|_2^2, \quad f_I(\alpha) \equiv \frac{1}{2}\|\bm A_I\bm \alpha - \bm b_I\|_2^2.
\end{align}
The $f_B$ objective encodes the first term in Eq.~\eqref{eq:winding_surface_optimization}, $f_K$ is Tikhonov regularization on the optimization variables, and $f_I$ encodes Eq.~\eqref{eq:Itarget} for avoiding the trivial solution $\bm \alpha = 0$.
The $\kappa$, $\sigma$, and $\lambda$ hyperparameters control the relative important of each loss term in the optimization.
This is equality-constrained sparse regression $-$ an optimization problem commonly appearing across science and plasma physics~\cite{kaptanoglu2023sparse}, for which a number of effective algorithms are available~\cite{bertsimas2020sparse,bertsimas2022backbone,bertsimas2022learning}. In fact, we have recently formulated stellarator magnet optimization using a large number of permanent magnets in a similar manner~\cite{kaptanoglu2022permanent,kaptanoglu2022greedy}. However, this is a challenging optimization problem in high-dimensions and with many constraints since the $l_0$ norm is nonconvex and nonsmooth. 

\begin{figure}
    \centering
    \includegraphics[width=\linewidth]{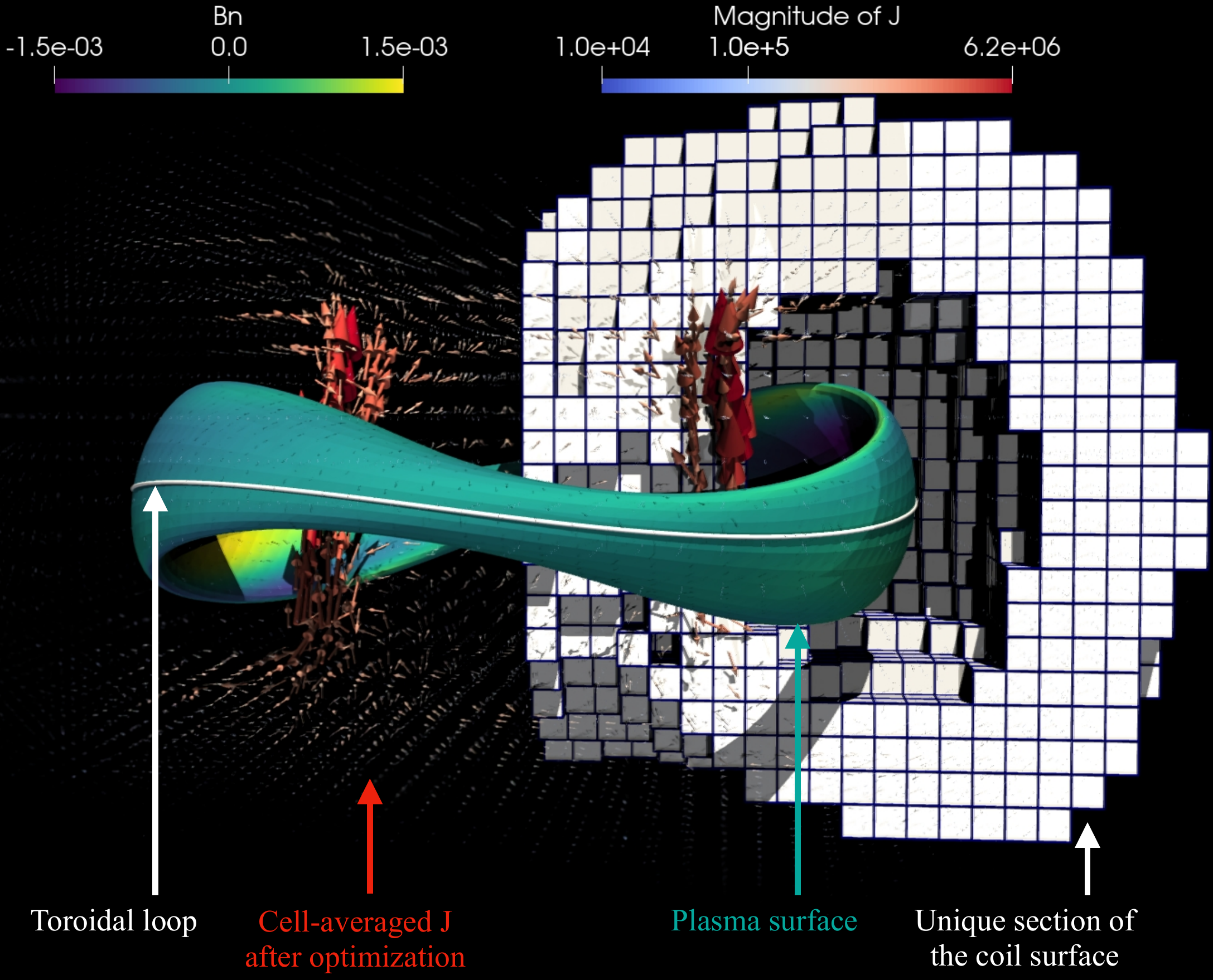}
    \caption{Full optimization geometry for the Landreman-Paul QA stellarator coil solution without any sparsity promotion ($\lambda = 0$). Only the unique quarter of the voxel grid is pictured. $\bm B \cdot \hat{\bm n}$ errors are shown on the plasma surface and the cell-averaged $\bm J$ solution vectors are color-coded by  $\|\bm J\|$.}
    \label{fig:setup}
\end{figure}

\begin{figure*}
    \centering
\includegraphics[width=\linewidth]{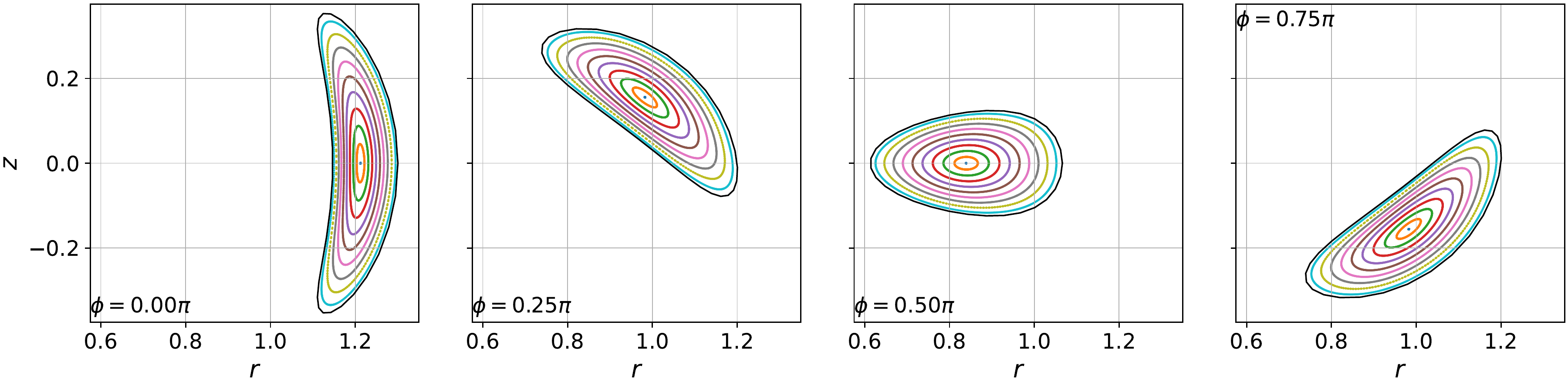}
    \caption{Poincaré plots showing the field produced by the voxels solution in Figure~\ref{fig:setup}, for the Landreman-Paul QA stellarator. Black lines indicate the plasma boundary $S$ that was targeted during optimization.}
    \label{fig:poincare_QA}
    \vspace{-0.2in}
\end{figure*}

For illustration of how the various optimization terms relate to the geometry, we show the full optimization geometry in Fig.~\ref{fig:setup} for the stellarator introduced in Sec.~\ref{sec:qa_design}. This includes the volume-averaged current density solution $\bm J(\bm r')$ in the voxels (indicated by the vectors), the unique part of the current voxel grid (the white cubic mesh), the toroidal loop $\gamma$ (white curve), and the plasma surface (mixed green colors, with $\bm B\cdot\hat{\bm n}$ errors plotted on the surface). The currents in the solution tend to be very strong on the inboard (small major radius) side where the plasma surface is vertically elongated. Poincaré plots in Fig.~\ref{fig:poincare_QA} illustrate that this current density solution reproduces the desired plasma equilibrium to high accuracy.


\subsection{Relax-and-split solution for topology optimization}\label{sec:relax_and_split}
High-dimensional, constrained, and nonconvex problems can be effectively solved with well-known algorithms, e.g., the Broyden–Fletcher–Goldfarb–Shanno (BFGS) algorithm~\cite{liu1989limited}, if the problem is smooth. However, the $l_0$ loss term is nonsmooth and therefore we need specialized algorithms. 

Relax-and-split methods, also called penalized decomposition methods, solve optimization problems by splitting them into two simpler subproblems. In the context of sparse regression, one set of optimization variables is used to solve the linear least-squares term and a set of convex constraints, and the second set is used to address the nonsmooth and/or nonconvex sparsity-promoting loss term~\cite{zheng2019unified,champion2020unified,kaptanoglu2021promoting}. Then a ``relaxation'' $L_2$ loss term is introduced to minimize the difference between the two sets of optimization variables. This approach for solving sparse regression problems has also been applied successfully to solve high-dimensional $l_0$-minimization problems arising in imaging science and compressive sensing; see, e.g., \cite{zhang2013l0,nikolova2010fast,attouch2010proximal}.

Mathematically, the relax-and-split method reformulates Eq.~\eqref{eq:final_opt} to become (taking $\sigma = 0$ and $\kappa = 0$ here for clarity):
\begin{align}\label{eq:split}
    &\min_{\bm \beta}\left\{\min_{\bm \alpha}\left\{\frac{\|\bm A\bm \alpha - \bm b\|_2^2}{2} + \frac{\|\bm \alpha - \bm \beta\|_2^2}{2\nu}\right\} + \lambda\|\bm \beta\|_0^G\right\},\\ \notag 
    &s.t. \quad \bm C\bm \alpha = \bm 0.
\end{align}
Notice there are now two optimization problems, one for $\bm \alpha$ and one for $\bm \beta$, and we can control how closely these variables match by tuning the $\nu$ hyperparameter.
The idea is now to iteratively solve this problem by variable projection $-$ fixing one variable while optimizing over the other $-$ and repeating until convergence is found. Consider initial conditions for the optimization variables, $\bm\alpha^{(0)}$ and $\bm\beta^{(0)}$.
The solutions in the $k$-th iteration we denote as $\alpha^{(k)}$ and $\beta^{(k)}$, so that,
\begin{align}
\label{eq:convex_solve}
   \bm \alpha^{(k)} \equiv \argmin_{\bm \alpha}&\left\{\frac{\|\bm A\bm \alpha - \bm b\|_2^2}{2} + \frac{\|\bm \alpha - \bm \beta^{(k-1)}\|_2^2}{2\nu}\right\},\\ \notag 
    s.t. \quad &\bm C\bm \alpha = \bm 0, \\ \label{eq:beta_solve}
    \bm \beta^{(k)} \equiv \argmin_{\bm \beta}&\left\{\frac{1}{2\nu}\|\bm \alpha^{(k)} - \bm \beta\|_2^2 + \lambda\|\bm \beta\|_0^G\right\}.
\end{align}
Problem~\eqref{eq:convex_solve} is a linear least-squares with affine constraints.
For most of the high-resolution results in the present work, the problem dimensions get large and $\bm A^T \bm A$ becomes very costly to compute and store in memory. 
Fortunately, iterative solvers are suitable for efficiently solving high-dimensional linear systems. 
In practice, we use the MINRES algorithm with an approximate Schur complement preconditioner~\cite{elman2014finite}, since it requires only matrix-vector products of the matrices appearing in Eq.~\eqref{eq:convex_solve}. Note that the algorithm and preconditioning can take advantage of the fact that $\bm C$ is a sparse matrix, since it encodes flux matching constraints for each cell's boundaries; only the cell and its (at most) six neighboring cells are involved in each of the constraints.
As long as the update to $\bm \alpha$ can be made reasonably computationally efficient, this relax-and-split strategy is effective for addressing the $l_0$-regularized topology optimization problem even if additional nonconvex terms are added. 
Next, the outer optimization problem~\eqref{eq:beta_solve} has a solution via the proximal operator,
    \begin{align}
    \bm \beta^{(k)} = prox_{\nu\lambda \|(.)\|^G_0}(\bm \alpha^{(k)}).
\end{align}
For the \textit{non-overlapping group} $l_0$ norm, the proximal operator is an analytic function akin to the traditional $l_0$, i.e., hard thresholding the norm of each subgroup. This process is repeated iteratively for $k$ iterations until some convergence criteria for $\bm \beta^{(k)}$ or $\bm \alpha^{(k)}$ is satisfied. Note that this algorithm is efficient for solving Eq.~\eqref{eq:final_opt} and any iterative algorithm for this problem relies on parallelizable matrix-vector products.
Finally, the full optimization in Eq.~\eqref{eq:split} can be solved many times for increasingly large values of $\lambda$, using the previous solution as an initial condition for the next optimization problem with larger $\lambda$. This process increasingly produces solutions that look like thin, high-current loops, i.e., realistic coils. 

Lastly, hyperparameter scans were performed and documented in Appendix~\ref{sec:appendix_hyperparameter_scans} to demonstrate convergence with respect to various geometrical quantities and find useful values of the optimization-related hyperparameters $\lambda$, $\sigma$, $\kappa$, and $\nu$. 

\subsection{Discrete symmetries in stellarators}
Symmetries play an important role for stellarators and discrete symmetries provide reductions in the required number of variables for performing coil optimization. Subsequently, most stellarators to date have been designed with discrete field-period and stellarator symmetries. Field-period symmetry refers to a periodicity in the magnetic field with respect to the number of periods, $n_p$ in a full toroidal turn. In cylindrical coordinates, taking $\zeta$ for the moment as the canonical azimuthal angle, 
\begin{align}
    \bm B(R, \zeta + 2\pi/n_p, Z) = \bm B(R, \zeta, Z).
\end{align}
Since $\bm B_\text{target}$ exhibits this property at the plasma surface $S$, and we desire that $\bm B_\text{coil}\cdot\hat{\bm n}$ matches $\bm B_\text{target}\cdot\hat{\bm n}$ on $S$, the coils should also exhibit field-period symmetry. In other words, we need only design coils for the unique $\zeta \in [0,  2\pi / n_p)$ part of the plasma surface and the unique $\zeta' \in [0, 2\pi / n_p)$ part of the current voxel grid. However, a simple Cartesian grid of voxels is used in the present work, which can only replicate this symmetry in $\bm B_\text{coil}$ if $n_p = 2$ or $4$, since otherwise the cubes cannot be stitched together properly. Therefore other values $n_p = 3, 5, 6,$ etc. must use a voxel grid defined in the entire $\zeta' \in [0, 2\pi)$. Fortunately, Appendix~\ref{sec:appendix_hyperparameter_scans} illustrates that our algorithm scales well with the number of voxels and therefore stellarators with these values of $n_p$ can still be readily optimized. Future work could address this geometrical issue by working with a cylindrical grid of voxels and associated basis functions, which would exhibit a continuous rotational symmetry. 

Next, a stellarator symmetric field is one in which, in a $(R, \zeta, Z)$ cylindrical coordinate system, $B_R$ is odd with respect to an inversion about the line $\zeta = 0$, $Z = 0,$ while $B_Z$ and $B_\zeta$ are even. This constraint on parity amounts to reducing the number of degrees of freedom by a factor of two. We can now design coils for the unique $0 \leq \zeta \leq \pi / n_p$ part of the plasma surface and the unique $0 \leq \zeta' \leq \pi / n_p$ part of the current voxel grid. For instance, for a stellarator that is two-field-period and stellarator symmetric, only a quarter of the plasma surface and a quarter of the current voxels are required. The contribution to the plasma surface from the remaining voxels is obtained not by optimization but by a set of rotations and parity flips akin to what is done in other coil optimization techniques for stellarators. Subsequently, the $\bm \alpha$ inherit the appropriate symmetries and the flux jump constraints are made consistent across the full voxel grid. 

\section{Results}\label{sec:results}
To demonstrate that our optimization problem can generate new coil designs, we consider three stellarators: the Landreman-Paul QA and QH configurations~\cite{landreman2022magnetic} as well as the recent two-field-period Goodman QI stellarator.~\cite{goodman2022constructing}. All three stellarators are scaled to 1 meter major radius and a plausible, laboratory-scale $B\approx 0.1$ T, averaged along the major radius. The exact values are not an important choice because these plasmas have no intrinsic length scale and subsequently solutions can always be appropriately rescaled.

Since the primary focus of this work was methodology and exploration of  coil topology, most of the results in the present work were generated with modest current voxel grid sizes.
Subsequently sufficient $f_B$ minimization is often not achieved in these examples, such that the achieved magnetic fields differ some from the target fields. Future work could ameliorate this issue by focusing on a particular stellarator design and performing high resolution runs, with more extensive variations of hyperparameters. We also address the voxel $f_B$ errors by using the current voxel solutions to initialize coil filaments, which are further optimized to low $f_B$ error and shown to reproduce the desired plasma. 

\begin{figure}
    \centering
    \includegraphics[width=0.9\linewidth]{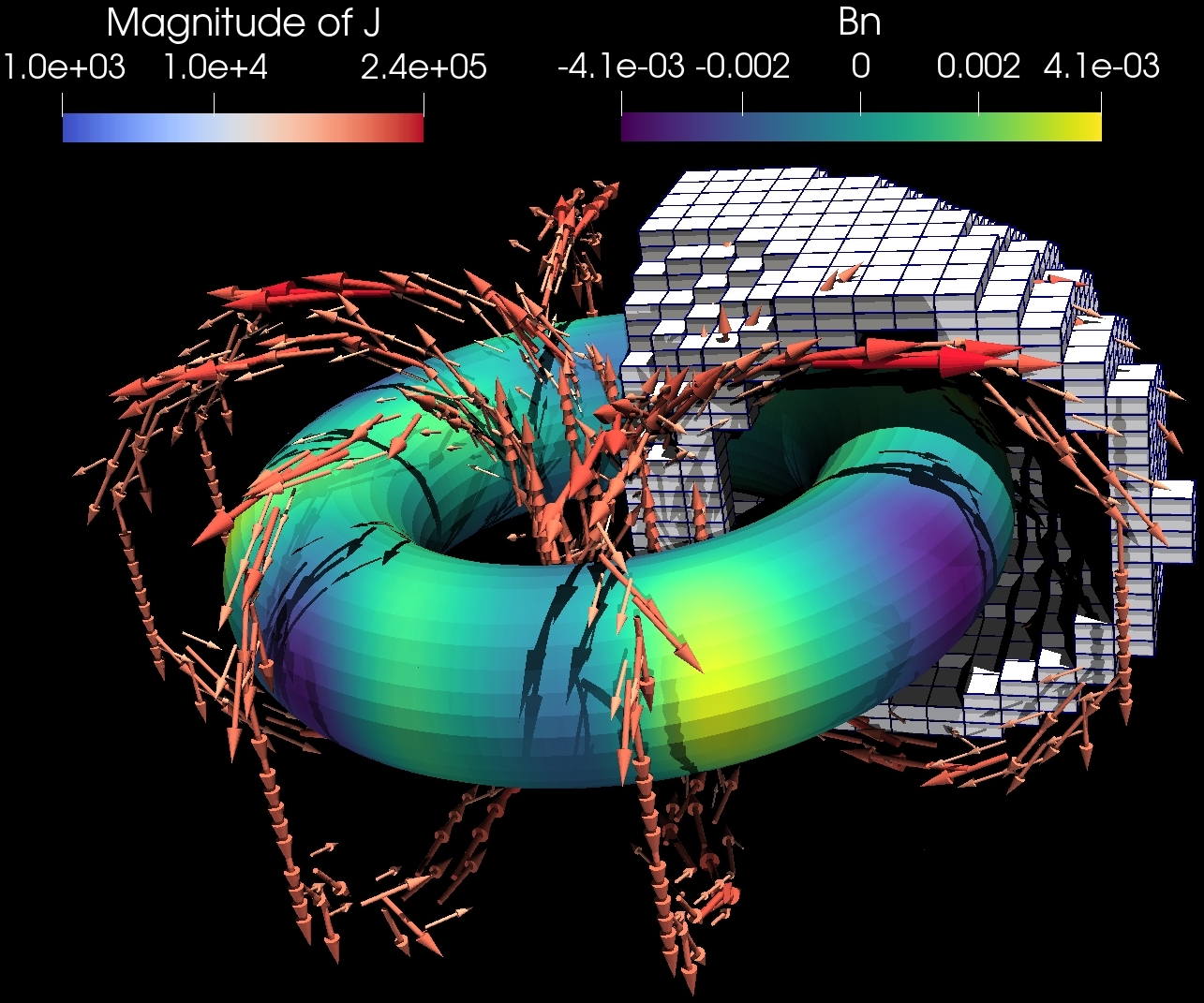}
    \caption{Optimization for an axisymmetric torus with a two-field-period and stellarator symmetric current voxel grid leads to two toroidal field coils in the unique part of the grid. $\bm B \cdot \hat{\bm n}$ errors are shown on the plasma surface and the cell-averaged $\bm J$ solution vectors are color-coded by $\|\bm J\|$.}
    \label{fig:torus}
\end{figure}

Before showing the stellarator designs, we test our method for an axisymmetric torus. 
We consider a case with no plasma current, so the axisymmetric target surface corresponds to a purely azimuthal target field.
The torus has continuous rotational symmetry, but the Cartesian voxel grid does not, so for convenience we prescribe that the coil volume is two-field-period and stellarator symmetric. A representative result is illustrated in Fig.~\ref{fig:torus}, with two coils per half field period, so there are eight coils in total. As expected for an axisymmetric azimuthal target field, we obtain approximately planar toroidal field coils.
The current density is not perfectly planar due to staircasing effects of the rectangular current voxels. 
This reasonable result for an axisymmetric target field provides initial evidence that our method is working properly and is capable of producing discrete coils.

\begin{figure}
    \centering
\includegraphics[width=0.83\linewidth]{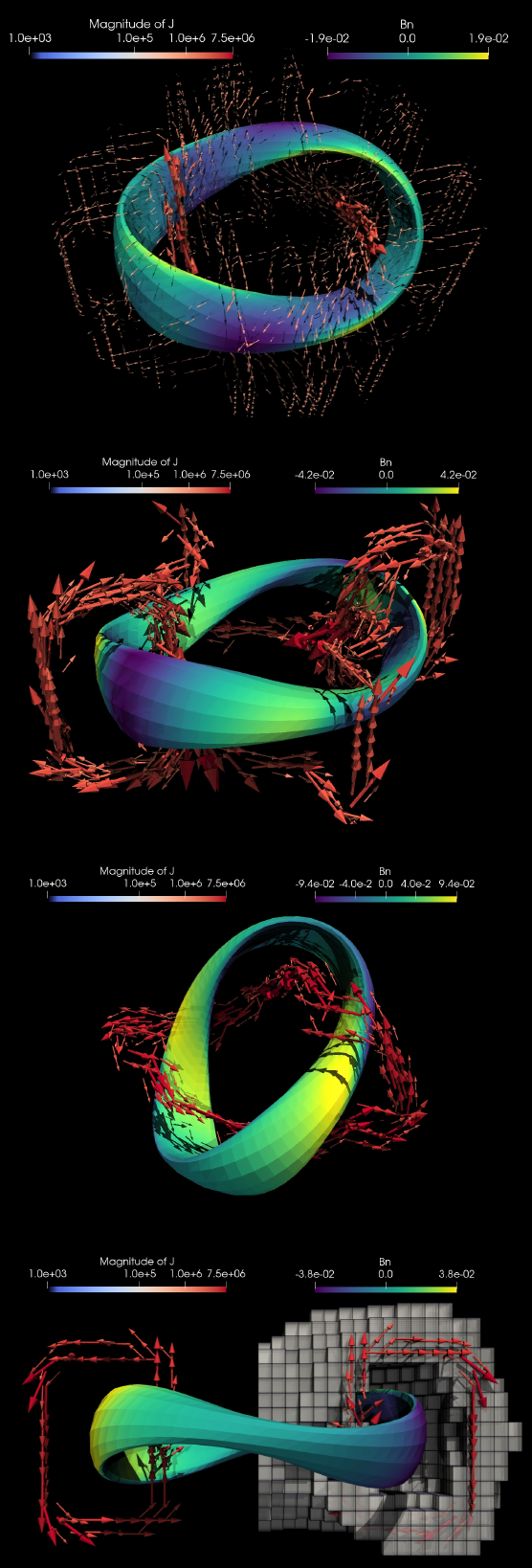}
    \caption{Landreman-Paul QA stellarator coil solutions of increasing sparsity from top to bottom. Only in the bottom illustration is the unique quarter of the voxel grid pictured. $\bm B \cdot \hat{\bm n}$ plasma surface errors are shown and the cell-averaged $\bm J$ solution is illustrated and color-coded by $\|\bm J\|$.}
    \label{fig:QA_solutions}
\end{figure}

\subsection{Coil designs for the Landreman-Paul QA stellarator}\label{sec:qa_design}
The Landreman and Paul QA stellarator~\cite{landreman2022magnetic} is stellarator-symmetric and two-field-period symmetric, meaning that only one quarter of the plasma surface and one quarter of the total coil optimization variables need to be determined. 
\begin{figure*}
    \centering
\includegraphics[width=\linewidth]{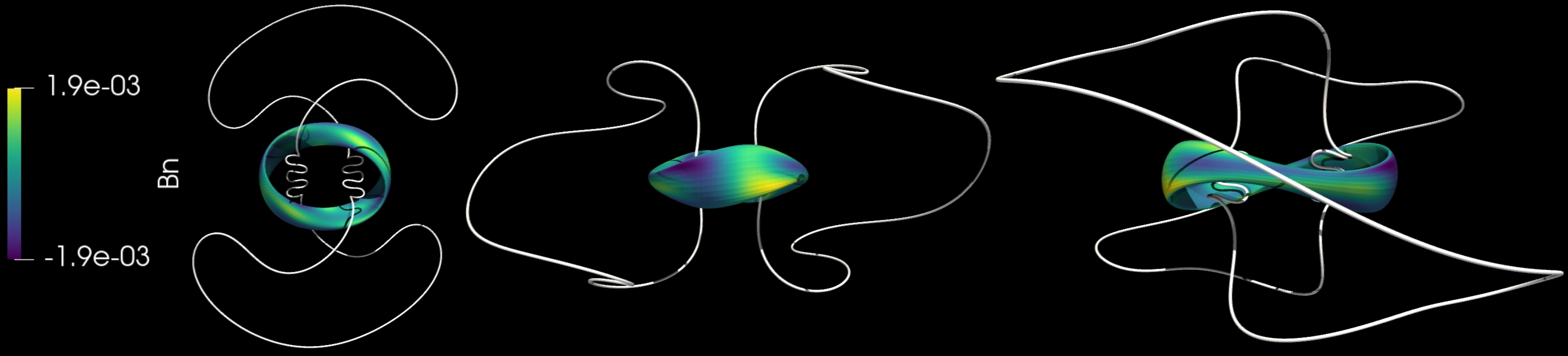}
    \caption{Three views of a 40 meter filament coil generated from a voxel solution for the Landreman-Paul QA stellarator.}
    \label{fig:QA_filament}
    \centering
\includegraphics[width=\linewidth]{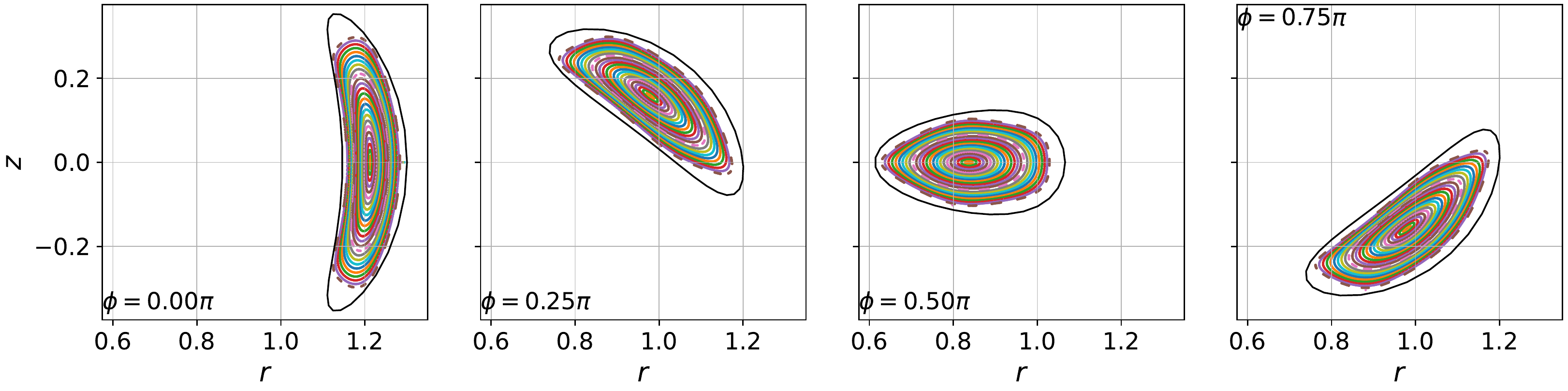}
    \caption{Poincaré plots for the figure-eight filament coil solution showing minor degradation of the plasma surfaces.}
    \label{fig:QA_filament_poincare}
\end{figure*}

Figure~\ref{fig:QA_solutions} illustrates the results for the QA stellarator at varying levels of sparsity-promotion. At each stage, unique topologies are exhibited, although almost all of the solutions have strong currents near the inboard (small major radius)  side of the plasma surface where it is vertically elongated, a result commonly observed in coil optimization for stellarators. Both modular and helical coil solutions are obtained. Some of these solutions are too complex for realistic engineering designs, but can be used as an initial condition for filamentary optimization. In particular, one of the most sparse solutions consists of a single figure-eight coil that links through the QA stellarator; an interesting topological choice for a helical coil. 

With demountable joints, a single optimized coil for the whole device could be an attractive design, so we now use this solution to initialize a filamentary coil optimization. Assuming that the Cartesian coordinates of the coil are unique when written as a function of the toroidal coordinate $\zeta$ or poloidal coordinate $\theta$, we can transform the identified curve $\bm x(\theta)$ to the Fourier basis used for filament optimization,
\begin{align}
    \label{eq:filament_basis}
    \bm x(\theta) = \frac{1}{2}x_{c, 0} + \sum_{m=1}^{M} \bm x_{c, m}\cos(m \theta) + \bm x_{s, m}\sin(m \theta).
\end{align}
The coefficients of the expansion are determined as usual by the orthogonality of the basis functions. However, voxel solutions typically have a finite thickness and may have nonzero neighbors. Moreover, the assumption of uniqueness with respect to $\zeta$ or $\theta$ implies a sufficient level of sparsity in the solution. In practice, if this uniqueness condition holds, we extract a curve from the unique $\zeta$ locations and apply a generous moving average to the Cartesian coordinates of the curve. This process results in small deviations from the original curve of voxels identified during optimization, but usefully eliminates the ambiguity in defining the curve and importantly retains the voxel topology. 
Once the curve, Fourier coefficients, and $I_\text{target}$ are specified, the filamentary optimization can be initialized and then performed. We omit the details, e.g. the hyperparameters and objective terms, of this optimization here but the methodology can be found in Zhu et al.~\cite{zhu2017new} and the results can be entirely reproduced in an example in the SIMSOPT code~\cite{SimsoptURL}.

The filament optimization results for the figure-eight coil in the third panel of Fig.~\ref{fig:QA_solutions} are illustrated in Fig.~\ref{fig:QA_filament}. This single, 40 meter long, helical coil is able to produce a solution accurate enough to generate good flux surfaces as illustrated in Fig.~\ref{fig:QA_filament_poincare}, though with a reduced volume compared to the original target configuration. 
For comparison, the Wechsung et al.~\cite{wechsung2022precise} coil set, with the same 1 meter major radius plasma and somewhat improved solution errors,
found sixteen modular coils (four unique coils) with total length of approximately 72 meters. One 40 meter long coil could be challenging to fabricate off-site and transport on-site, but with demountable joints the coil can be fabricated and transported in separate pieces. 
Moreover, it is challenging to further improve the plasma surfaces here without making the already long coil significantly longer. It is common knowledge that saddle coils are needed to assist the helical coil, but further exploration on this point is beyond the scope of this paper.

Despite these caveats, there are well-known benefits to using helical coils for stellarators, and in particular a single helical coil is attractive from a diagnostic access and engineering standpoint. Helical coils also minimize toroidal ripple, the small-scale errors that arise from modular coils with small coil-plasma distance. Notably, compared to the modular coil solution, our figure-eight solution has larger coil-plasma separation, larger coil-coil separation (except for the small regions in the center with high curvature), and more room for diagnostic access and neutron-absorbing blankets (required for nuclear fusion reactors). 
\begin{figure}
    \centering
\includegraphics[width=0.855\linewidth]{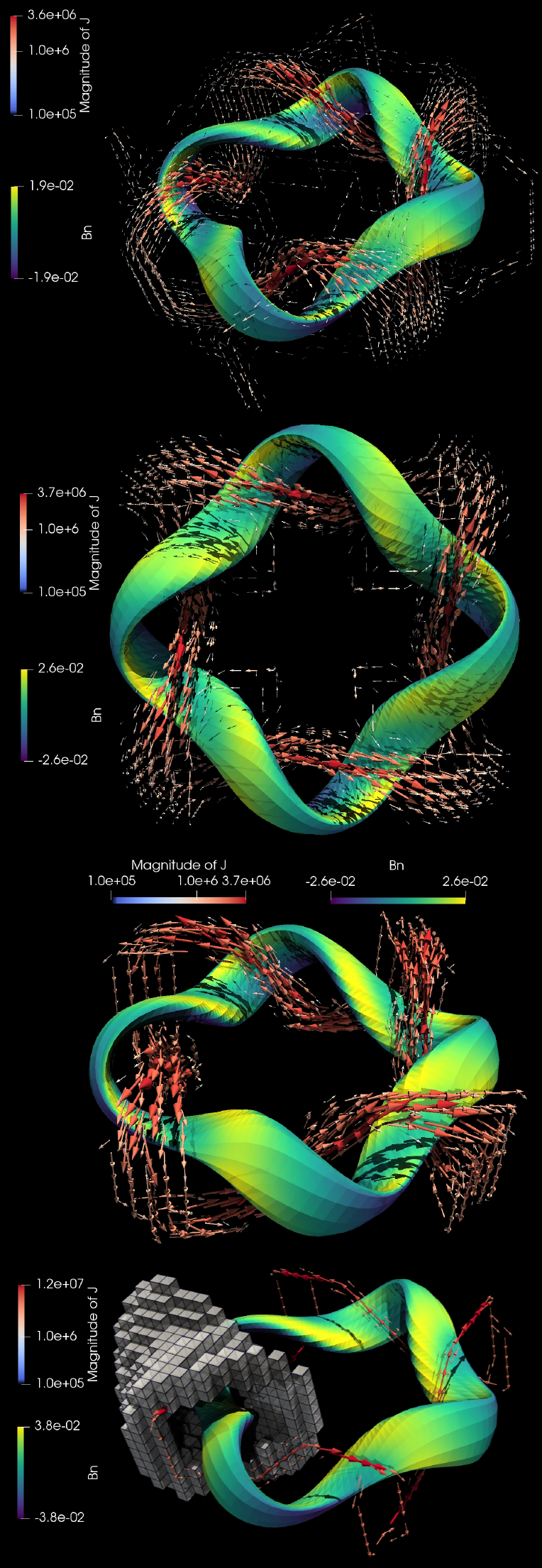}
    \caption{Landreman-Paul QH stellarator coil solutions of increasing sparsity from top to bottom. Only in the bottom illustration is the unique eighth of the voxel grid pictured. $\bm B \cdot \hat{\bm n}$ plasma surface errors are shown and the cell-averaged $\bm J$ solution is illustrated and color-coded by $\|\bm J\|$. }
    \label{fig:QH_solutions}
\end{figure}

\subsection{Coil design for the Landreman-Paul QH stellarator}\label{sec:qh_design}
The Landreman and Paul QH stellarator~\cite{landreman2022magnetic} is stellarator-symmetric and four-field-period symmetric, so only one-eighth of the plasma and voxel grid is required for optimization. 
Figure~\ref{fig:QH_solutions} illustrates some of the exotic configurations found through optimization. As in many stellarator coil solutions, the currents tend to congregate near the inboard side of bends where reducing the normal magnetic field is challenging.
Moreover, the sparsest solution looks like a rectangular, four-field-period coil that is quadruple-linked with the stellarator. In fact, all of the solutions exhibit this underlying structure in the currents. Similar helical-coil solutions for other optimized stellarators have recently been investigated independently \cite{yamaguchi2019quasi, yamaguchi2021optimization}, and it is exciting to find a similar coil topology through a new optimization method. 

\begin{figure*}
    \centering
\includegraphics[width=\linewidth]{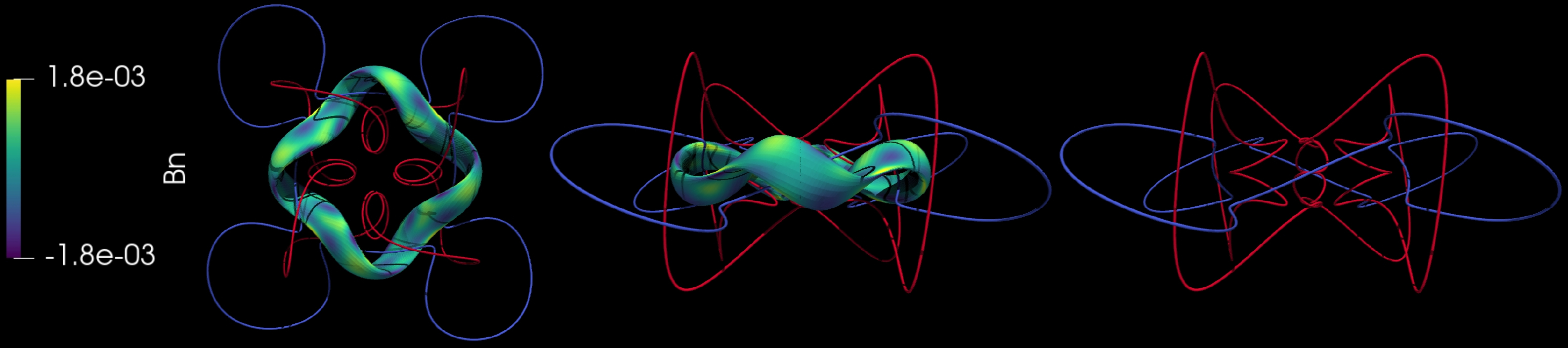}
    \caption{Three views of the helical coils with combined $\approx 24 + 29 = 53$ meter length, generated from a voxel solution for the Landreman-Paul QH stellarator. The right-most panel only shows the coils for a better understanding of the geometry.}
    \label{fig:QH_filament}
    \centering
\includegraphics[width=\linewidth]{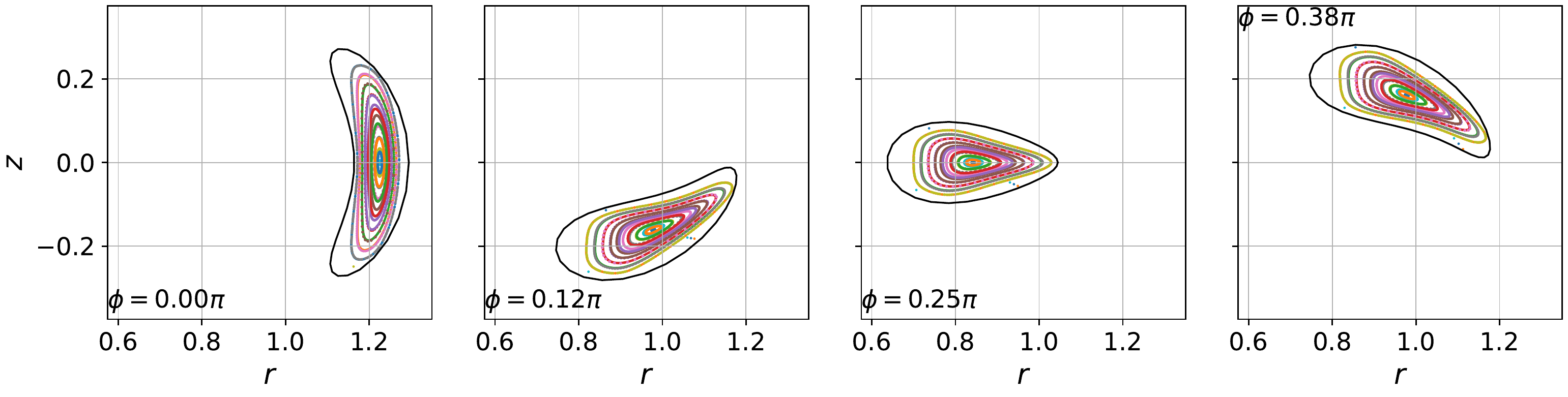}
    \caption{Poincaré plots for the coil set in Fig.~\ref{fig:QH_filament} showing minor degradation of the plasma surfaces.}
    \label{fig:QH_filament_poincare}
\end{figure*}

Like the previous example, we use the four-field-period current voxel solution coil in the third panel of Fig.~\ref{fig:QH_solutions} to initialize a filament optimization. With a single coil, we were unable to sufficiently reduce $\bm B \cdot \hat{\bm n}$ enough to accurately produce the desired plasma surfaces. Instead, we initialize two coils with the same topology, with one coil slightly perturbed in space from the voxel solution. The resulting filament optimization with these two helical coils is similar in spirit to the optimizations in Yamaguchi et al.~\cite{yamaguchi2021optimization} and Elder et al.~\cite{elder2022helical}, which both utilize multiple helical coils. A two-coil filament solution, with combined length of 53 m, is illustrated in Fig.~\ref{fig:QH_filament}. The coils are accurate enough to produce flux surfaces in the Poincaré plots in Fig.~\ref{fig:QH_filament_poincare}, though with some distortions compared to the original configuration. Despite the coil complexity, these helical coils could be a useful alternative to the modular coils typically used for four-field period stellarators. For instance, the four-field-period and stellarator symmetric HSX device has 48 coils (6 unique coils), with total coil length $\sim 90$ m~\cite{anderson1995helically, almagri1999helically}, and subsequently a neutron-absorbing blanket is infeasible and diagnostic access is limited. Lastly, as far as we aware, the pair of intertwined helical coils in Fig.~\ref{fig:QH_filament} represents the first successful coil set for this QH stellarator in the literature. 

\begin{figure}
    \centering
\includegraphics[width=0.95\linewidth]{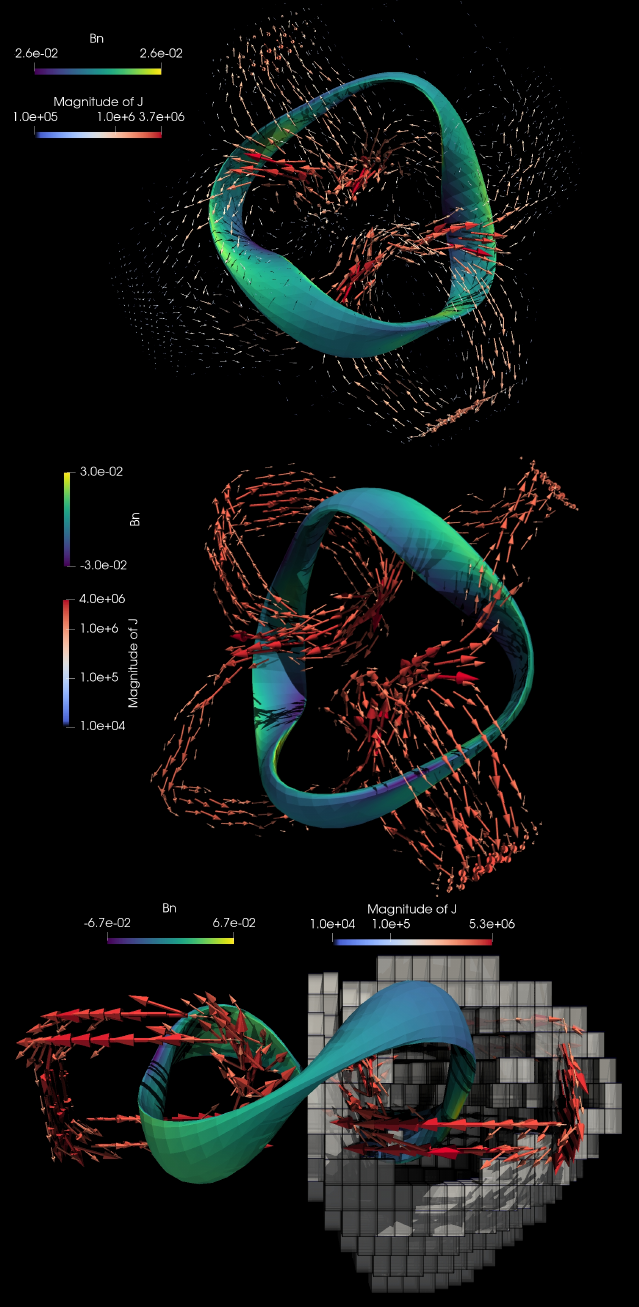}
    \caption{Goodman QI stellarator coil solutions of increasing sparsity from top to bottom. The bottom illustrates the unique quarter of the voxel grid. $\bm B \cdot \hat{\bm n}$ surface errors are shown and the cell-averaged $\bm J$ is illustrated and color-coded by $\|\bm J\|$.}
    \label{fig:QI_solutions}
\end{figure}

\subsection{Coil design for the Goodman QI stellarator}\label{sec:qi_design}
To conclude the results, we use our new method to compute some initial coils for the two-field-period and stellarator symmetric QI stellarator found in Goodman et al.~\cite{goodman2022constructing}.
Figure~\ref{fig:QI_solutions} illustrates some of the optimized coil configurations. Interestingly, the currents tend to be important near the straight, ``race-track''  parts of the plasma surface, and it seems to be challenging to find coils that are spatially distributed around the plasma. The sparsest solution consists of a single rotated window-pane coil (two identical coils after symmetrizing) and figure-eight helical coils and more complex topology also appear. Initializing a figure-eight helical coil or tilted modular coil from this solution could be interesting future work.

\section{Conclusion}\label{sec:conclusion}
We have formulated topology optimization based on sparse regression and the $l_0$ norm, and additionally provided an algorithm that can effectively solve a large subclass of topology optimization problems across scientific disciplines. 
To demonstrate our method, we have designed a new approach for inverse magnetostatics, computing topologically-unconstrained electromagnets.
While stellarator coils were considered as a specific application here,
we expect the method can be applicable to other areas in which a target magnetic field must be produced, such as magnetic resonance imaging or particle accelerator optics.
Additionally, we have provided examples of several exotic topological solutions for three different stellarators of interest to the plasma physics community. 

This method is new and subsequently there is ample room for improvement and refinement. Future work includes: implementation of higher-order basis functions, tetrahedral meshes, algorithmic speedups through improved iterative solvers and preconditioners or improved sparse regression algorithms, additional loss terms in the optimization for reducing coil forces or coil curvatures, reformulation as stochastic optimization to control for coil errors, and much more. A reformulation may be possible that builds in the current conservation by construction, rather than as constraints in the optimization problem.

This method explores a high-dimensional nonconvex optimization space that may exhibit more exotic or more useful solutions than the ones found in this initial work. Although not explored in this work, initial conditions for the optimization can bias the solutions towards producing a particular topological structure or a certain number of identifiable coils. Along with adding additional engineering-related optimization terms, clever initial conditions could facilitate real-world coil designs; some of the coil solutions in this work are presented because they are interesting topologically, but these solutions could present serious engineering challenges. It may require such initial conditions or additional loss terms in order to fully reproduce the types of solutions found using filament optimization. Indeed, this work is perhaps most compelling for providing principled topology choices to initialize more complex filament optimization for stellarators. 

\section{Acknowledgements}\label{sec:acknowledgements}
Thanks to Georg Stadler for optimization improvements and paper suggestions. We would like to acknowledge assistance from Todd Elder, Elizabeth Paul, Alan Goodman, Stefan Buller, and many others in the Simons Collaboration on Hidden Symmetries and Fusion Energy. 
This work was supported by the U.S. Department of Energy under award DEFG0293ER54197 and through a grant from the Simons Foundation under award 560651. This research used resources of the National Energy Research Scientific Computing Center (NERSC), a U.S. Department of Energy Office of Science User Facility located at Lawrence Berkeley National Laboratory, operated under Contract No. DE-AC02-05CH11231.

\appendix

\section{Defining the current voxel grid}\label{sec:appendix_grid_generation}
As is often done in other magneto-static optimization problems for stellarators, e.g., permanent magnet optimization, we define the permissible volume for voxels as the space between two toroidal limiting surfaces. A simple transformation can be used to generate this volume. We begin by initializing a uniform Cartesian grid, incorporating the discrete symmetries of the plasma surface if possible, in a large region surrounding the plasma. The plasma boundary surface is extended outward by a constant multiple of the unit normal to generate an inner toroidal boundary. An outer limiting surface is generated similarly, using the normal vectors on the inner toroidal surface. For moderately shaped equilibria, these simple transformations work well to generate a toroidal volume. Any of the original grid cells that are not between the inner and outer surfaces are eliminated with a ray-tracing routine.
It is straightforward to extend this method for more complex grids. For instance, diagnostic ports can easily be included by removing any intersecting grid cells, and updating the flux jump constraints accordingly for the remaining grid cells.

\section{Matrix forms of the loss terms}\label{sec:appendix_formalism}
In this section, we show how the various optimization objectives appearing in stellarator coil optimization can be formulated as linear terms in the optimization variables $\bm \alpha$.
We take advantage of possible stellarator and field-period symmetries by using $n_\theta n_\zeta$ quadrature points in poloidal and toroidal angles $(\theta, \zeta)$ on the plasma surface to write for any scalar surface quantity Q:
\begin{align}
    \int_S Qd^2r &= \sum_{i=1}^{n_p}\int_{0}^{2\pi}d\theta\int_{0}^{2\pi/n_p}d\zeta nQ \\ \notag &\approx \sum_{i=1}^{n_p}\sum_{j=1}^{n_\theta n_\zeta}\Delta\theta_{j}\Delta\zeta_{j}n_jQ_j.
\end{align}
Here $n_p$ is the value of the field-period symmetry, $\bm n = \frac{\bm{dr}}{\bm d\theta} \times\frac{\bm{dr}}{\bm d\zeta}$ are the surface normal vectors, and $n = \|\bm n\|$.
Plugging in the basis expansion for the $\bm J_k$ in each cell, the coil contributions at each quadrature point $\bm r_j$ can be summarized as,
\begin{align}
\label{eq:Bnormal_coil_matrix}
    \bm B_\text{coil}(\bm r_j)\cdot\bm n_j
    &= -\frac{\mu_0}{4 \pi}\sum_{k=1}^{D}\bm\alpha\cdot\int_{V_k'}\frac{\hat{\bm n}\times (\bm r_j - \bm r_k')}{\|\bm r_j - \bm r_k'\|^3}\cdot\bm\phi(\bm r_k') d\bm r_k', \\ \notag
     &= \sum_{i=1}^N\alpha_iG_{ij} = \bm G\cdot\bm\alpha, \\ \notag 
     \bm G &\equiv -\frac{\mu_0}{4 \pi}\int_{V_k'}\frac{\hat{\bm n}\times (\bm r_j - \bm r_k')}{\|\bm r_j - \bm r_k'\|^3}\cdot\bm\phi(\bm r_k') d\bm r_k'.
\end{align}
The total inductance matrix $\bm G$ can be computed only once before optimization begins and the integrals over $V_k'$ are evaluated with a tensor-product quadrature grid.
The loss term associated with the normal magnetic field on the plasma boundary becomes
\begin{align}
\label{eq:Bnormal_plasma_matrix}
    f_{B}(\bm \alpha) &\equiv \frac{1}{2}\left\|\bm A\bm \alpha - \bm b\right\|^2, 
    \\ \notag 
    b_j &\equiv \sqrt{\Delta\theta_{j}\Delta\zeta_{j}\|\bm N_j\|_2}\bm B_{\text{target}, j}\cdot\bm n_j, \\ \notag 
    A_{ji} &\equiv\sqrt{\Delta\theta_{j}\Delta\zeta_{j}\|\bm N_j\|_2}G_{ji},
\end{align}
where $\Delta\theta_{j}$ and $\Delta\zeta_{j}$ indicate the grid spacing in the two angular directions.
Equation~\eqref{eq:Bnormal_plasma_matrix} is linear least-squares in the $\bm \alpha$ optimization variables, as desired. 

There is a similar term that comes from the requirement
\begin{align}\label{eq:Itarget_append}
    \mu_0 I_\text{target} = \oint_\gamma(\bm B_\text{coil} - \bm B_0) \cdot \bm{dl},
\end{align}
defined earlier in Eq.~\eqref{eq:Itarget} to avoid the trivial solution. Note first that 
\begin{align}
\label{eq:Bcoil_Itarget}
    \bm B_\text{coil}(\bm r)\cdot \bm{dl}
    &= -\frac{\mu_0}{4 \pi}\sum_{k=1}^{D}\bm\alpha\cdot\int_{V_k'}\frac{\bm {dl}\times (\bm r - \bm r_k')}{\|\bm r - \bm r_k'\|^3}\cdot\bm\phi(\bm r_k') d\bm r_k', \\ 
    &= \bm F \cdot\bm\alpha,
\end{align}
where $\bm F$ is the equivalent to $\bm G$ but with the replacement $\hat{\bm n} \to \bm{dl}$. Then in total we have
\begin{align}
    \sum_{j=1}^{n_\gamma}D_{ji} \alpha_i - e_j &= \mu_0I_\text{target}, \\ \notag 
    D_{ji} &\equiv \Delta\zeta_j F_{ji}, \quad
    e_j \equiv\Delta\zeta_j\bm B_{0, j} \cdot \bm dl_j.
\end{align}
Now add a row of zeros to $\bm D$ and append $\mu_0I_\text{target}$ to the end of $\bm e$. 
Then Eq.~\eqref{eq:Itarget_append} can be written in the form
\begin{align}\label{eq:Itarget_form}
   \bm A_I\bm \alpha = \bm b_I,  \quad \bm A_I &\equiv \sum_{j=1}^{n_\gamma + 1}D_{ji}, \quad \bm b_I \equiv \sum_{j=1}^{n_\gamma + 1} e_j,
\end{align}
which we are free to recast as a loss term to be minimized in the optimization, 
\begin{align}\label{eq:Itarget_lstsq}
   f_I(\bm \alpha) \equiv \frac{1}{2}\|\bm A_I\bm\alpha - \bm b_I\|_2^2.
\end{align}
The flux jump condition in Eq.~\eqref{eq:flux_jumps_vanish} also needs to be written in terms of the $\bm \alpha$ optimization variables:
\begin{align}
\int_{V_k' \cap V_l'} &\hat{\bm n}'\cdot\left[\bm J_k(\bm r_k') - \bm J_l(\bm r_k')\right]d^2 r_k' = 0
\\ \notag
&= \sum_{i=1}^N\int_{V_k' \cap V_l'} \hat{\bm n}'\cdot\left(\alpha_{ik}\bm \phi_{ik} - \alpha_{il}\bm \phi_{il}\right)d^2 r_k',
\end{align}
where the $l$ index denotes the index of the adjacent cell. Many of the cells will have fewer than six constraints because of duplicates from other cells, i.e., two adjacent cells need only a single constraint for their mutual interface. Stacking the constraints from all the cells produces
\begin{align}\label{eq:constraint_final}
    C_{ki} \alpha_i =  0, \quad
    C_{ki} \equiv\int_{V_j' \cap V_l'} \hat{\bm n}'\cdot\left(\bm \phi_{ij} - \bm \phi_{il}\right)d^2 r_j',
\end{align}
with an appropriate index mapping between $k$ and $(j, l)$.
We have now defined the equality constraints required for the current density to match flux jumps at cell interfaces. Since the current densities are divergence-free within cells, this additional constraint produces globally divergence-free current density.

There is a subtlety present in the constraints in Eq.~\eqref{eq:constraint_final}. The $\bm C$ matrix is not full rank and this appears to due to the limited expressiveness of the linear finite element basis to represent the current density in each cell. In practice, this is only a potential issue for preconditioning, or computing $\bm C^{-1}$ via the pseudoinverse. Alternatively, this problem could be somewhat ameliorated by the use of higher-order polynomial basis functions.

Lastly, we can add Tikhonov regularization, 
\begin{align}\label{eq:tikhonov_appendix}
    f_K(\alpha) \equiv \frac{1}{2D}\|\bm \alpha\|_2^2,
\end{align}
with a factor of $D^{-1}$ introduced to compensate for  the dependence of $\|\bm \alpha\|_2^2$ on the number of voxels. 
Finally, the complete optimization problem is
\begin{align}
    \min_{\bm \alpha}&\left\{f_B(\alpha) + \kappa f_K(\alpha) + \sigma f_I(\alpha) + \lambda\|\bm \alpha\|_0^G\right\},\\ \notag 
    s.t. \quad &\bm C\bm \alpha = \bm 0.
\end{align}
Tikhonov regularization tends to be critical when $\lambda = 0$, especially for the MINRES preconditioning, but less important when $\lambda \neq 0$ since the group-sparsity term tends to regularize the solution anyways.

\begin{table*}[t]
\centering
\begin{tabular}{ |p{2.5cm}|p{2cm}|p{10cm}|p{2cm}|  }
 \hline
 Hyperparameter & Type &
 Description & Default value  \\
 \hline
 $\lambda$ & Optimization &
 Specifies the strength of group sparsity-promotion. & 0  \\
 \hline
 $\nu$ & Optimization & How closely the $\bm\alpha^*$ and $\bm \beta^*$ solutions of Eq.~\eqref{eq:split} should match in $L_2$. & $\infty$\\
 \hline
 $\kappa$ 
& Optimization & Degree of Tikhonov regularization. & $10^{-15}$\\
 \hline
 $\sigma$ 
& Optimization & How stringently to match the prescribed $I_\text{target}$ through a toroidal loop. & 1\\
 \hline
 $D$ 
& Geometric & Number of grid cells. & $\sim 10^3-10^5$\\
 \hline
 $N'$ 
& Geometric & Number of points used for each cell's Biot-Savart calculations. & $6^3$ \\
 \hline
 $n_\zeta n_\theta$ 
& Geometric & Number of uniformly-spaced quadrature points on the plasma surface. & $16^2$ \\
 \hline
 $n_\gamma$ 
& Geometric & Number of uniformly-spaced quadrature points on the toroidal loop. & 8\\
 \hline
\end{tabular}
\caption{Description of the hyperparameters for our proposed coil optimization. With reasonable values for the convex optimization, $\lambda = 0$, $\nu \to \infty$, $\sigma = 1$, and $\kappa = 10^{-15}$, the geometric parameters have converged by $D \approx 10,000$, $N_x \approx 6$, $n_\zeta n_\theta = 64^2$, and $n_\gamma = 8$. 
We find that these values are fairly robust to different stellarator configurations.}
\label{tab:hyperparameters}
\end{table*}

\section{Hyperparameter scans}\label{sec:appendix_hyperparameter_scans}
Here, we investigate the convergence of the algorithm solutions with respect to the geometric hyperparameters, using the Landreman and Paul QA stellarator~\cite{landreman2022magnetic}. 
A description of each of the hyperparameters is shown in Table~\ref{tab:hyperparameters}. It was found that convergence of the geometric hyperparameters was essentially independent of the particular stellarator configuration. 

There are four primary algorithm hyperparameters: $\kappa$ controlling the amount of Tikhonov regularization, $\sigma$ controlling how closely to match the prescribed total current through a toroidal loop, $\nu$ controlling the amount of relaxation between $\bm \alpha$ and $\bm \beta$, and $\lambda$ controlling the amount of group sparsity. In the convex limit, without sparsity promotion, only $\kappa$ and $\sigma$ are relevant. In the $\sigma \to 0$ or $\kappa \to \infty$ regimes, the optimization correctly arrives at the trivial solution. 

\begin{figure}
    \centering
    \includegraphics[width=0.95\linewidth]{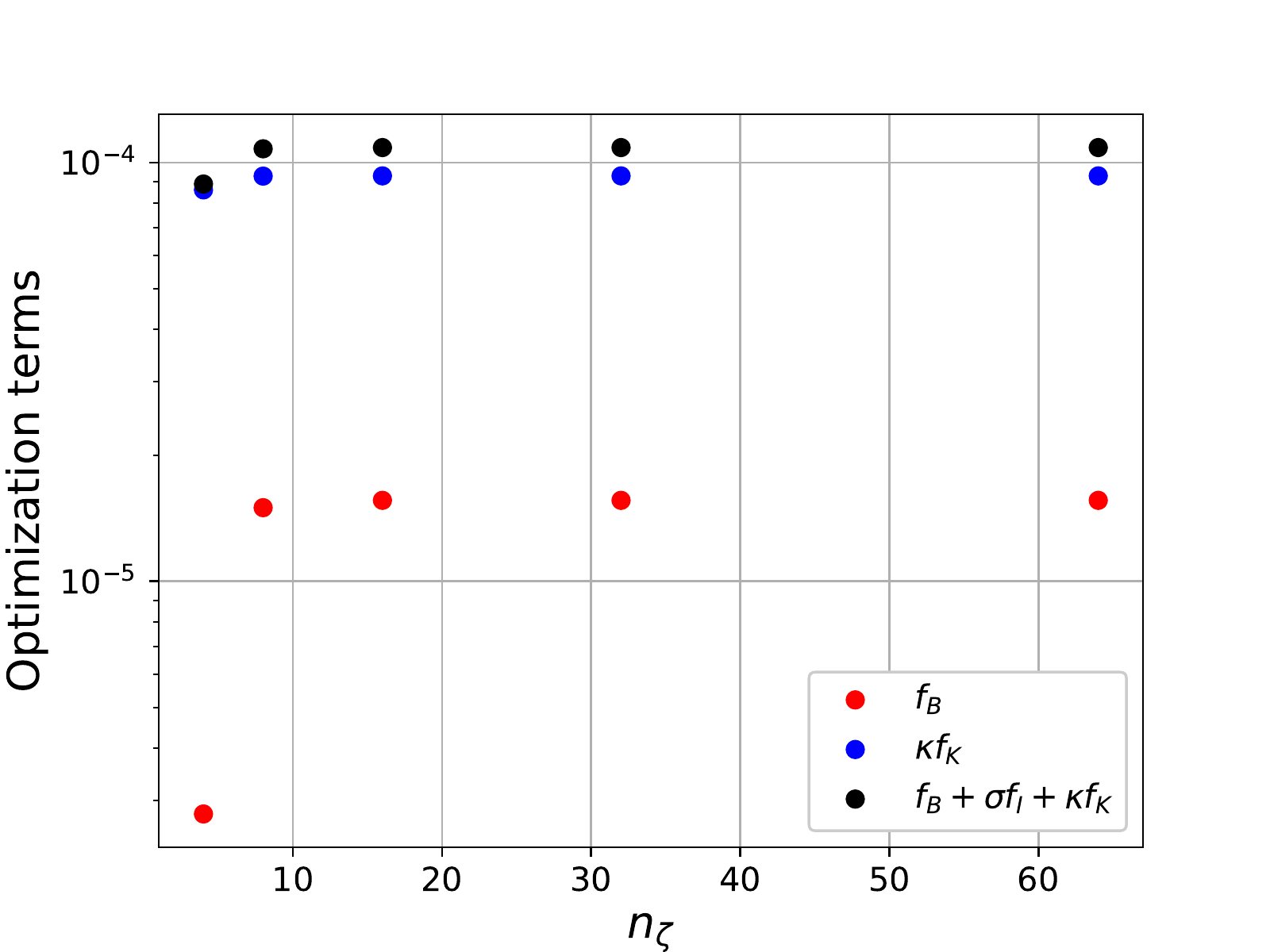}
    \caption{Optimization results showing convergence with respect to the number of uniformly-sampled quadrature points on the plasma surface, $n_\zeta n_\theta = n_\zeta^2$.}
    \label{fig:convergence_nphi}
\end{figure}

There are a number of geometric quantities in the optimization: 
the spatial resolution of $V'$ or equivalently the number of unique grid cells $D$, the number of points $N'$ used for intra-cell integrations of the Biot Savart law, and the number of quadrature points $n_\zeta n_\theta$ used for the plasma boundary and $n_\gamma$ for the toroidal loop. Note that $n_\zeta n_\theta$ denotes the number quadrature points on the half-field-period surface, so that the total number of quadrature points for this stellarator is $4n_\zeta n_\theta$ (and similarly for $n_\gamma$ and $D$). It was found that $n_\gamma = 8$ is already sufficient for optimization, since the exact shape of the toroidal loop is anyways unimportant for our purposes, and therefore we omit it from the more careful convergence studies described below. Convergence with respect to $n_\theta n_\zeta$ is illustrated in Fig.~\ref{fig:convergence_nphi}; $n_\theta = n_\zeta = 16$ is already well-converged. 

\begin{figure}
    \centering
\includegraphics[width=0.9\linewidth]{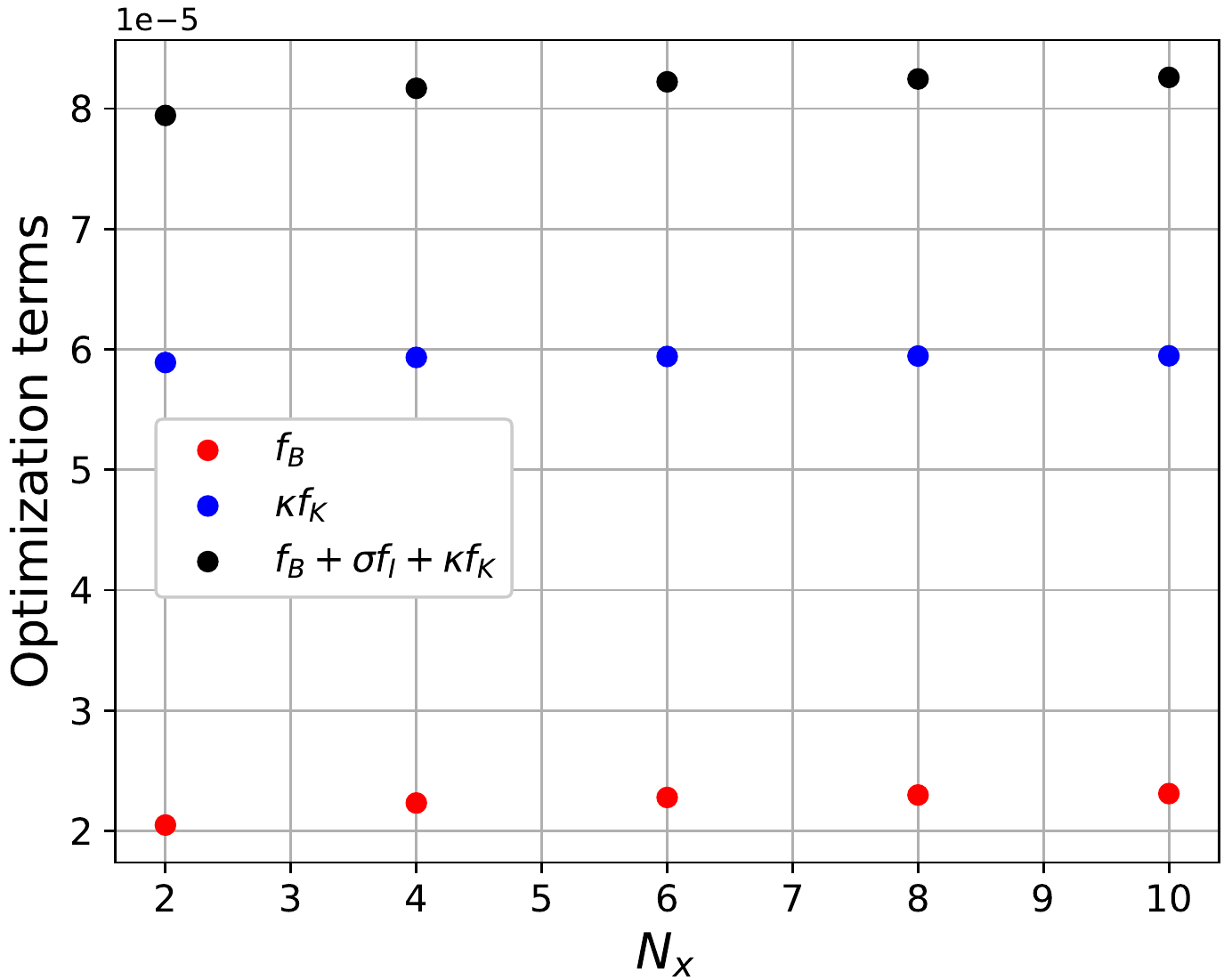}
    \caption{Optimization results showing convergence with respect to the number of Biot-Savart integration points per grid cell, $N' = N_x^3$. 
    }
    \label{fig:convergence_Ngrid}
\end{figure}

For convergence studies we take cubic cells, $N' = N_x^3$, and determine the minimal $N_x$ value for accurate Biot-Savart calculations from each cell. For our purposes, ``convergence'' refers to the convergence of the solution found in the convex limit of the optimization problem, $\lambda = 0$ and $\nu \to \infty$. We consider typical optimization hyperparameters $\sigma = 1$ and $\kappa = 10^{-15}$ for a reasonably well-posed optimization problem with $f_B \sim \kappa f_K$. Then we start with $N_x = 1$ and increase this value until these increases provide no change in the final solution to the optimization problem. The Biot-Savart calculation is then calculated accurately enough to at least produce the same global minimum. Figure~\ref{fig:convergence_Ngrid} illustrates that $N_x \approx 6$ is sufficient for the Biot-Savart calculations to be accurate enough that the optimization is converged.

\begin{figure}
    \centering
    \includegraphics[width=0.975\linewidth]{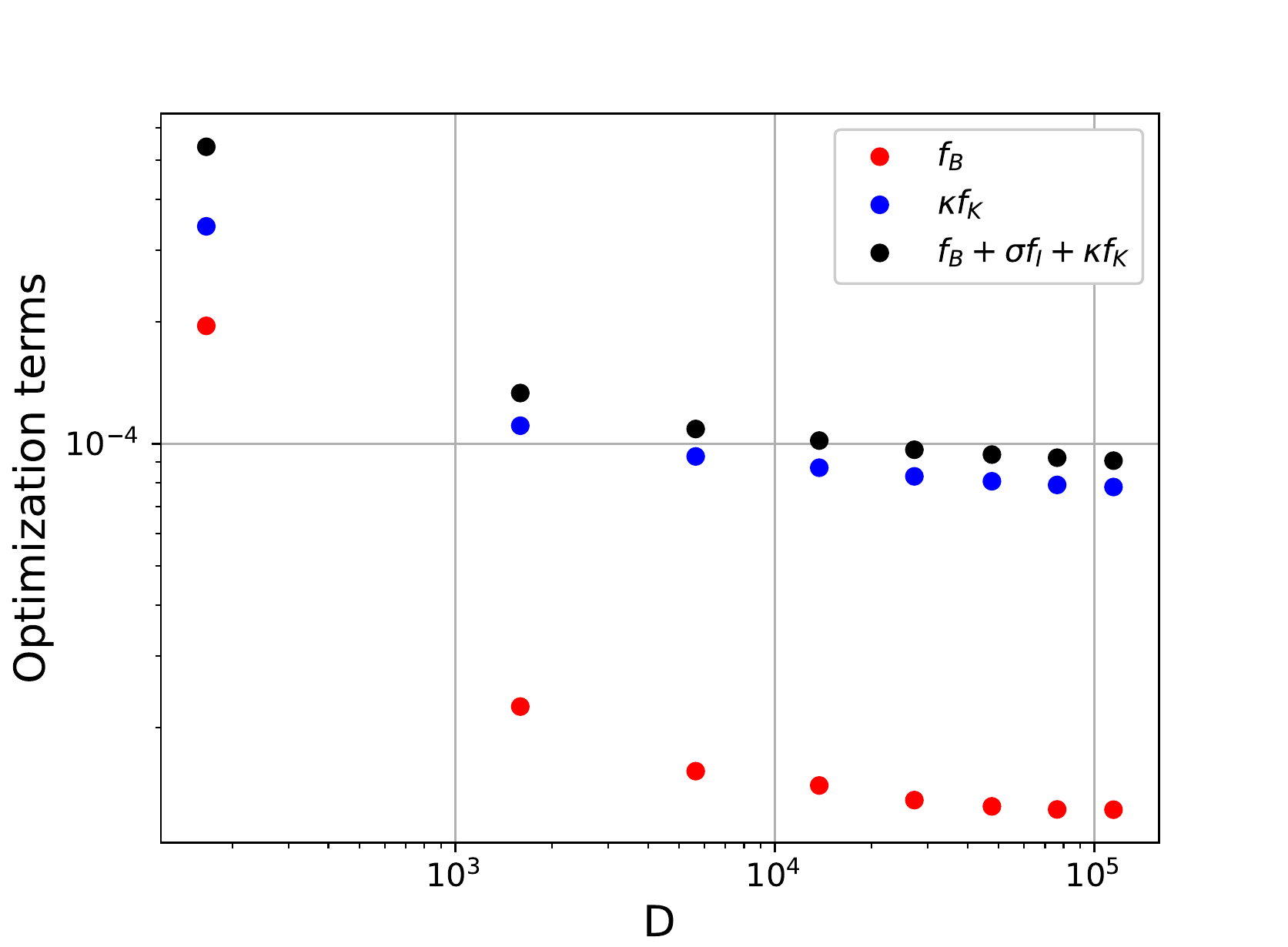}
    \caption{Optimization results showing  convergence with respect to the number of grid cells, D, while holding the total current voxel volume constant. }
    \label{fig:convergence_D}
\end{figure}

\begin{figure}
    \centering
    \includegraphics[width=\linewidth]{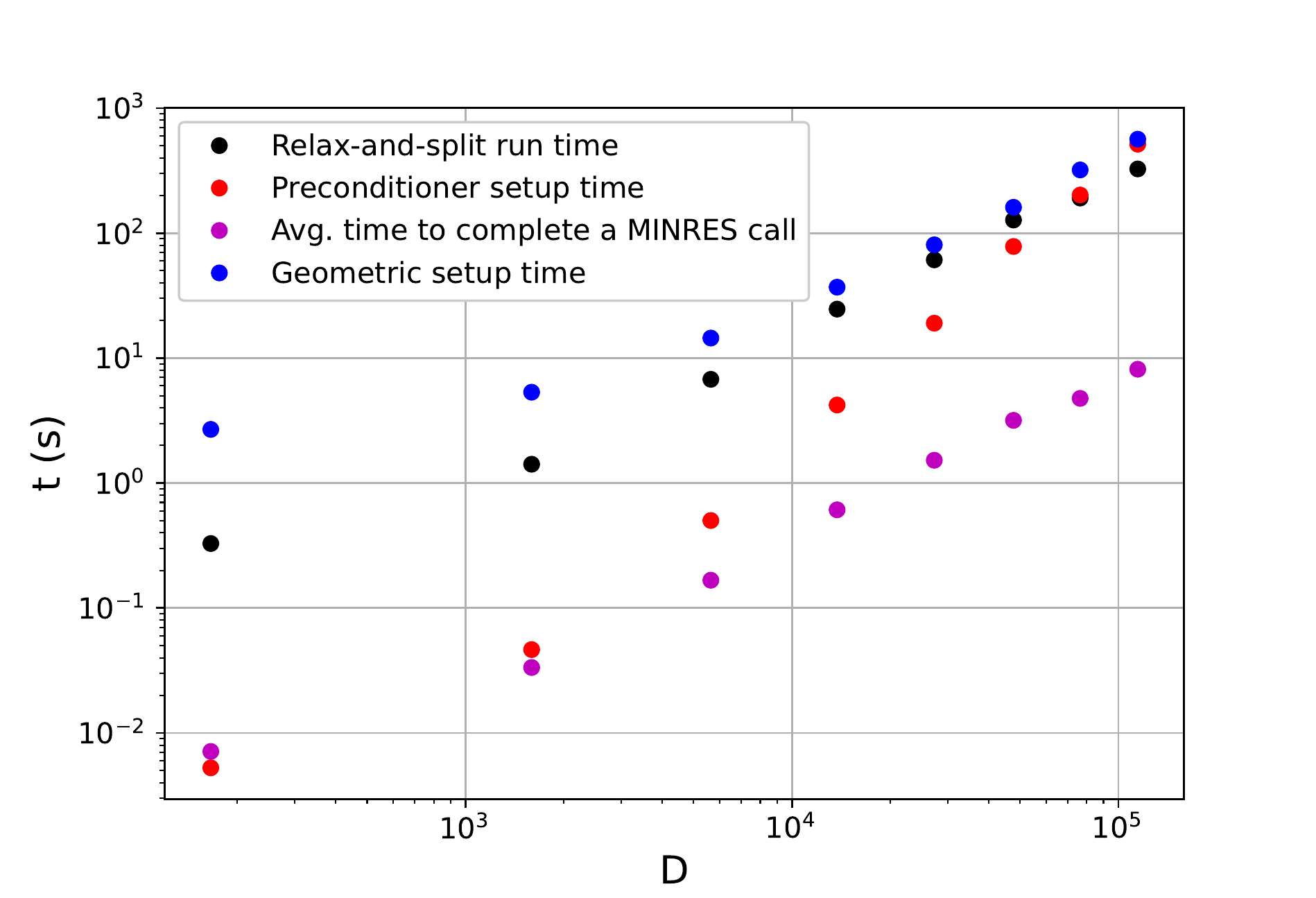}
    \caption{Scaling of the relax-and-split computational time and other important elapsed times with the number of grid cells, $D$. Note that geometric and preconditioner setup are one-time calculations before optimization begins.  }
    \label{fig:timing}
\end{figure}

Similarly, we increase the number of voxels, $D$, by keeping the overall grid volume constant, while increasing the number of cells. The cells subsequently get smaller and smaller and therefore provide a test for convergence. If $\kappa = 0$, this is an ill-posed problem that in general can continue to find better global minima as more coil degrees of freedom are added to the problem. With large enough $\kappa$, the problem is well-posed and Fig.~\ref{fig:convergence_D} illustrates convergence with respect to $D$. Notice that the Tikhonov loss term in Eq.~\eqref{eq:tikhonov_appendix} is scaled by the number of voxels.

Furthermore, we test the scaling between various computational times and the number of voxels in Fig.~\ref{fig:timing}. The code is parallelized via Openmp and xsimd~\cite{xsimd}. All runs used a single AMD EPYC 7763 CPU on the Perlmutter supercomputer, with 64 cores per CPU. Additional specifications for these nodes are available online.

The algorithmic MINRES scaling with $D$ is  favorable; from $D \sim 10^3 \to D \sim 10^5$, the time for a complete preconditioned MINRES solution only increases by an order of magnitude. The computational time for a full relax and split solve is calculated with a fixed $\lambda = 10^5$, $\nu = 10^{14}$. The many calls to MINRES are the bottleneck in the overall optimization.
Therefore the time for a relax and split solve scales similarly (here we use 40 iterations of relax-and-split and therefore 40 calls to MINRES). The geometric and preconditioning setup scalings are somewhat less favorable but importantly these quantities need only be computed once before optimization begins. 
Lastly, note that the right-most points represent a solution using $114,208$ (unique) grid cells and therefore $571,040$ optimization parameters in $\bm \alpha$. In this case, the matrix $\bm A^T\bm A$ is dense with $\sim 326$ billion nonzero elements.
 \bibliography{finite_build_coils}

\end{document}